\documentclass[prd,amsmath,floats,amssymb, floatfix,superscriptaddress,nofootinbib,onecolumn]{revtex4} 

\usepackage[plainpages=false, colorlinks=true, anchorcolor=blue, linkcolor=blue, citecolor=blue, bookmarks=false]{hyperref}
\usepackage[T1]{fontenc}
\usepackage{graphicx}% Include figure files
\usepackage{dcolumn}% Align table columns on decimal point
\usepackage{bm}% bold math
\usepackage{longtable}
\usepackage{adjustbox}

\usepackage{float}
\usepackage[caption=false]{subfig}
\usepackage[paperwidth=210mm,paperheight=297mm,centering,hmargin=2cm,vmargin=2.5cm]{geometry}

\newcommand{\rthis}[1]{\textcolor{black}{#1}}

\begin{document}
\include{notations}
\preprint{APS/123-QED}

\title{A test for the redshift dependence of $\sigma_8$ using $f\sigma_8$ measurements}% Force line breaks with \\
%\thanks{A footnote to the article title}%

\author{Siddhant Manna}
 \altaffiliation{Email:ph22resch11006@iith.ac.in}%Lines break automatically or can be forced with \\
\author{Shantanu Desai}
 \altaffiliation{Email:shntn05@gmail.com}
\affiliation{
 Department of Physics, IIT Hyderabad Kandi, Telangana 502284,  India}

%\collaboration{MUSO Collaboration}%\noaffiliation

%\author{Charlie Author}
% \homepage{http://www.Second.institution.edu/~Charlie.Author}
%\affiliation{
 %Second institution and/or address\\
 %This line break forced% with \\
%}%
%\affiliation{
 %}%
%\author{Delta Author}
%\affiliation{%
 %Authors' institution and/or address\\
 %This line break forced with \textbackslash\textbackslash
%}%

%\collaboration{CLEO Collaboration}%\noaffiliation

%\date{\today}% It is always \today, today,
             %  but any date may be explicitly specified

\begin{abstract}
We search for a redshift dependence of $\sigma_8^0$ using \rthis{23} growth rate measurements from redshift space distortions and peculiar velocity measurements as a consistency check of $\Lambda$CDM.  For this purpose we use the   dataset from ~\cite{Sagredo} consisting of 22 measurements, which has been vetted for internal consistencies. \rthis{Eighteen of these measurements have been obtained  from the Gold-2017 sample, collated from  multiple redshift space distortion surveys between 2009 and 2016, whereas the remaining four have been obtained from the eBOSS DR14 quasar survey. We also added one additional data point from the BOSS DR12 CMASS galaxy sample.}  We find that for this dataset the  $\sigma_8^0$  values are consistent between the low redshift and high redshift samples using three different redshift cuts.
This implies that the growth rate measurements are consistent with a constant $\sigma_8^0$ in accord with $\Lambda$CDM,  \rthis{assuming  there are no uncontrolled systematics in this dataset.}

%When we parameterize the redshift dependence of $\sigma_8^0$ as a linear function of redshift, we find that the data redshift-dependent term is consistent with zero  to within $1.2\sigma$. This implies that the growth rate measurements are consistent with a constant $\sigma_8^0$ in accord with $\Lambda$CDM.

\end{abstract}


\maketitle
%\tableofcontents
\section{\label{sec:level1}Introduction\protect}

The current concordance model ($\Lambda$CDM) consisting of 70\% dark energy and 25\% cold dark matter is the prevailing cosmological model for the universe. It is based on the Friedmann–Lemaître–Robertson–Walker (FLRW) metric and assumes that the universe is spatially flat, homogeneous, and isotropic ~\cite{Ratra07}.  The model has been very successful in explaining a wide range of cosmological observations, including the cosmic microwave background (CMB), the abundance of light elements, and the large-scale structure of the universe~\cite{Planck2020}. 
%The Hubble constant, denoted by $H_0$, is a fundamental parameter in cosmology that represents the current expansion rate of the universe. It is typically measured in units of kilometres per second per megaparsec (km/s/Mpc). The value of $H_0$ is crucial for determining the age of the universe, the expansion history, and the composition of the cosmos.
%In the Lambda-CDM $\Lambda$CDM cosmological model, the prevailing model for the universe's evolution, $H_0$, is estimated to be around 67.4 km/s/Mpc. This value is based on measurements of the cosmic microwave background (CMB) radiation, a relic of the early universe.

Despite the overall success of the concordance $\Lambda$CDM model of cosmology, over the last few years, there   are still a few data-driven tensions with observations.
The most vexing problem is the intriguing discrepancy between the Hubble constant ($H_0$) values obtained from local measurements and those inferred from CMB data. Local measurements of $H_0$, primarily obtained from observing Cepheid variable stars and supernovae of Type Ia, typically yield a higher value,  around 73 km/s/Mpc compared to the CMB inferred value of around 67 km/s/Mpc. This discrepancy, known as the Hubble tension, poses a challenge to the $\Lambda$CDM model and has spurred a search for potential explanations ~\cite{Lahav,DiValentino,Verde,Riess22,Bethapudi,Freedman2021,Smoot23,Verde23,Vagnozzi23}.
Another vexing problem is the  $\sigma_8$/$S_8$ tension.
Here, $\sigma_8$  is a measure of the variance  of the matter density fluctuations on a scale of 8 $h^{-1}$ Mpc and at redshift $z=0$~\cite{Huterer22}. For the rest of the manuscript we use $\sigma_8^0$ to refer to $\sigma_8$ inferred at redshift of 0 to distinguish between growth rate measurements, which depend on $\sigma_8$ inferred at non-zero redshifts.  One can then define a weighted amplitude of matter fluctuations ($S_8$) by combining it with matter density ($\Omega_m$) and $\sigma_8^0$ as follows:
\begin{equation}
S_8 \equiv \sigma_8^0 \sqrt{\frac{\Omega_m}{0.3}} 
\end{equation}
Recent measurements of $S_8$ from different cosmological probes such as CMB and galaxy survey measurements have shown that $S_8$ inferred from Planck has a higher value of $S_8$ as compared to that obtained from galaxy surveys~\cite{Planck2020, Abdalla22,Abbott2020,Abbott2022, Asgari2021,Heymans2013,Joudaki2017,Hikage2019,Esposito2022,ACT23}. This is known as the $S_8$ tension.
An up-to-date summary of all recent tensions of the standard Cosmological model with observations is reviewed in ~\cite{Periv,Abdalla22,Peebles22,Banik}.

Concurrent with these tensions, there have been recent results, which have shown that some of the above cosmological constants depend on the redshift of the probes used to determine these constants. In other words they have been found to be  redshift-dependent.
For instance, some works have suggested that the Hubble constant might not be constant but could instead evolve with redshift, decreasing with increasing redshift \cite{Colgain2022,Colgain2022b,Wong2020,Krishnan2020,Malekjani2023,Dainotti2021,Hu2022,Jia2023,Millon2020}. In a similar vein,  there have been claims that the matter density parameter $\Omega_m$, increases with effective redshift~\cite{Colgain2022,Chuang2013,Malekjani2023,Risaliti2019,Yang2020,Khadka2020,Khadka2021,Khadka2022,Pourojaghi2022,Pasten2023}. 
%Most recently the  DES-Supernova~\cite{DES24}   and Union3~\cite{Union3} results on $\Omega_M$ are $1.5\sigma$ to $2.6\sigma$ discrepant with Planck. 

Some of these works have also found an anti-correlation between the variation of $\Omega_m$ and $H_0$.
Such a variation of the above cosmological constants with redshift implies  a  breakdown of the $\Lambda$CDM model assuming there are no uncontrolled systematic errors, since $\sigma_8^0$ and $H_0$ are integration constants within the $\Lambda$CDM model for an FLRW metric and do not evolve with redshift by definition~\cite{Wang1998}. Therefore, if their inferred values using probes at different redshifts vary, it would signal a breakdown of $\Lambda$CDM.

If $\Omega_m$ also varies  as a function of redshift,  this implies that the observed increase in $S_8$ must be driven by an increase in $\sigma_8^0$. Therefore, in order to test this {\it ansatz}, ~\citet{Adil2023} (A23, hereafter) searched for a redshift dependence of $S_8$ within $\Lambda$CDM and assuming a constant $\Omega_m$. For this purpose, they used $f\sigma_8$ measurements  from peculiar velocity and redshift space distortion measurements, where $f=\frac{d\ln \delta}{d \ln a}$~\cite{Nesseris}. This work  checked for a redshift dependence  of   $S_8$ after  bifurcating the dataset into two samples using the redshift cut of $z=0.7$. These measurements are agnostic to galaxy bias and hence provide a more robust probe. Using these measurements,  A23 showed that $S_8$ increases with redshift.

In this work, we carry out a slight variant of the analysis implemented in A23 using an independent dataset, where we search for a  redshift dependent $\sigma_8^0$ by dividing the samples according to redshift using three redshift cuts. Since by definition, $\sigma_8^0$ should not depend on the redshift of the probes used for its measurement, a statistically significant difference between its estimated value between the different redshift probes  would provide a hint of  breakdown of  $\Lambda$CDM or point to some unknown systematics in the data.
This manuscript is structured as follows. In Sect.~\ref{sec:level2}, we summarize the results of A23. In Sect.~\ref{sec:level3} we describe the data used, along with the analysis. Finally, our conclusions can be found in Sect.~\ref{sec:conclusions}. For our analysis, we assume a flat $\Lambda$CDM cosmology with $h=0.7$.

\section{Summary of A23}
\label{sec:level2}
A23 performed a consistency test of the Planck-$\Lambda$CDM cosmology to investigate the redshift evolution of the parameter $S_8$. They first subjected the analysis to 20 growth rate data points from peculiar velocity and redshift-space distortion (RSD) data~\cite{Nguyen}, and as a cross-check also redid  their analysis with 66 growth rate data points obtained from ~\cite{Kazantzidis2018} for a more comprehensive analysis.  
A23 imposed a prior on $\Omega_m$=$0.3111 \pm 0.0056$, which is informed by observational data from both the Planck CMB and BAO experiments~\cite{Planck2020}.
In their study, they employed a generalized matter density parameter, denoted by $\Omega(z)$~\cite{Wang1998}. It quantifies the relative density of matter in the universe compared to a critical density. It is defined  according to:
\begin{equation}
\Omega(z) := \frac{\Omega_m (z)}{{H(z)^2}/{H_0^2}} = \frac{\Omega_m (1 + z)^3}{1 - \Omega_m + \Omega_m (1 + z)^3}.    
\end{equation}
The corresponding  equation for $f\sigma_8(z)$ can be expressed as:
\begin{equation}
f\sigma_8 (z) = \sigma_8^0 \Omega^{6/11} (z) \exp \left( -\int_0^{z} \frac{\Omega^{6/11}(z')}{1 + z'}\right).dz'
\label{eq:fsigma8}
\end{equation}
The expression in Eq.~\ref{eq:fsigma8} is a valid approximation for $\Lambda$CDM~\cite{Wang1998} and the maximum difference with the exact expression (involving hypergeometric functions)~\cite{Nesseris} is about 1\% at $z =0$~\cite{Adil2023}. Since our aim is to test for a redshift dependence, we use the aforementioned equation, which would be accurate enough for our purpose.
A Bayesian regression analysis using three free parameters ($\Omega_m$,$\sigma_8$, and $S_8$)  was carried out using Eq.~\ref{eq:fsigma8} after bifurcating the $f\sigma_8$ measurements into a low redshift and high redshift sample with the boundary at $z=0.7$. Using 20 measurements, A23 found that for the low redshift data, $\Omega_m$ and $\sigma_8$ were consistent with Planck 2020, but $S_8$ was discrepant at about  $3.4\sigma$. For the high redshift data, the tension with the Planck 2020 measurements for all the three parameters is between $1-2\sigma$.  For the 66 growth measurements, the data was bifurcated at $z=1.1$. The $S_8$ values between low and high redshifts differ by about $2.8\sigma$.
Also,  the $S_8$ value below $z=1.1$ is in tension with the Planck value at about $5\sigma$.
%The analysis reveals a tension (discrepancy) in the $S_8$ parameter when comparing constraints from low redshift data with those from the Planck CMB data at lower redshifts. This tension is quantified as a $3\sigma$ deviation, indicating a statistically significant difference. Remarkably, the tension in $S_8$ with Planck at lower redshifts becomes consistent with Planck within $1\sigma$ at high redshifts. This suggests that the discrepancy observed at lower redshifts diminishes at higher redshifts. When the CMB+BAO $\Omega_m$ prior is absent, the analysis finds that shifts improve tensions ($>3\sigma$) with Planck in low redshift data in the parameters in high redshift data. This suggests that including the $\Omega_m$ prior plays a crucial role in reconciling tensions between different datasets. For 66 growth rate data points they report an overestimated $2.8\sigma$ tension between low redshift and high redshift data. It is noteworthy that the $S_8$ parameter derived from data below z = 1.1 exhibits a significant $~5\sigma$ tension with Planck predictions, while data above z = 1.1 is consistent with Planck at $~1\sigma$, emphasizing a redshift-dependent discrepancy in cosmological constraints. 

\section{Data Selection and Analysis}
\label{sec:level3}
The datset in A23 was  based on the compilations in  Refs~\cite{Kazantzidis2018,Perivolaropoulos2022}.
%which in turn  used  measurements of $f\sigma_8(z)$  from multiple sources~\cite{Beutler2012,Song2009,Blake2013,Blake2012B,Howlett2017,Okumura2016,Davis2011,Hudson2012,Turnbull2012,Tojeiro2012,Samushia2012,GilMarin2018,delaTorre2013,delaTorre17,Feix2017,Zhao2019,Hou2018,Shi2018,Alnchez2014,Chuang2016,Alam2017,Beutler2017,Wilson2016,Feix2015,Mohammad18,Chuang2013,Wang2018,Hawken2017,Howlett2015,Bautista2021,deMattia2021,Neveux2020}. In addition, we also carry out the analysis with the same set 20 measurements used in the main body of A23, originally used for the analysis in ~\cite{Nguyen2023} which is a subset of the  66 measurements dataset. The 20 measurements are independent of each other, whereas correlations could exist in the 66 measurements dataset. More details on the measurements can be found in A23 or in the aforementioned references.
However, the measurements of $f\sigma_8(z)$ can be prone to various systematic effects, necessitating careful consideration when selecting the sample. Since the datasets used in A23 could exhibit correlations among themselves, we use the dataset in ~\cite{Sagredo} which consists of 22 growth rate measurements (cf. Table I of ~\cite{Sagredo}). 
This dataset has been extensively scrutinized for systematics and internal consistency checks using the Bayesian approach prescribed in ~\cite{Amendola}. Eighteen data points in this sample are   based on the Gold-2017 compilation, consisting of independent  measurements of $f\sigma_8$ collated from different redshift space distortion based surveys \rthis{between 2009 and 2016}~\cite{Nesseris}. \rthis{These surveys include 6dFGS,  IRAS, 2MASS, SDSS, GAMA, SDSS-LRG-200, SDSS-CMASS, BOSS-LOWZ, WiggleZ, Vipers, FastSound. These measurements depend on the fiducial cosmology assumed by the surveys for the analysis. These were corrected by rescaling the growth rate measurements by the ratio of $H(z)D_a(z)$ of the cosmology to the fiducial one. More details on this rescaling can be found in ~\cite{Nesseris}.}
The remaining  four measurements were obtained from ~\cite{Zhao2019}, \rthis{obtained from SDSS-IV  eBOSS DR14 quasar survey~\cite{Dawson}.} To this dataset, we added a  recent measurement of $f\sigma_8 = 0.462 \pm 0.020$  at $z \approx 0.525$ obtained from clustering analysis of the BOSS DR12 CMASS galaxy sample~\cite{Enrique}.

%~\cite{Adil2023,Beutler2012,Blake2013,Blake2012B,Howlett2015,Okumura2016,Boruah2020,Turner2023,Pezzotta2017,Alam2021,Huterer2017}.They have also been compiled in~\cite{Nguyen2023}.
%~\cite{Adil2023,Beutler2012,Blake2013,Blake2012B,Howlett2015,Okumura2016,Boruah2020,Turner2023,Pezzotta2017,Alam2021,Huterer2017}.
%Including more recent growth rate data not only increases the size of our dataset but also allows us to capture the latest observational insights into the evolution of $f\sigma_8(z)$across different redshifts. 

\begin{figure*}[!t]
  \centering
  \includegraphics[width=0.8\textwidth]{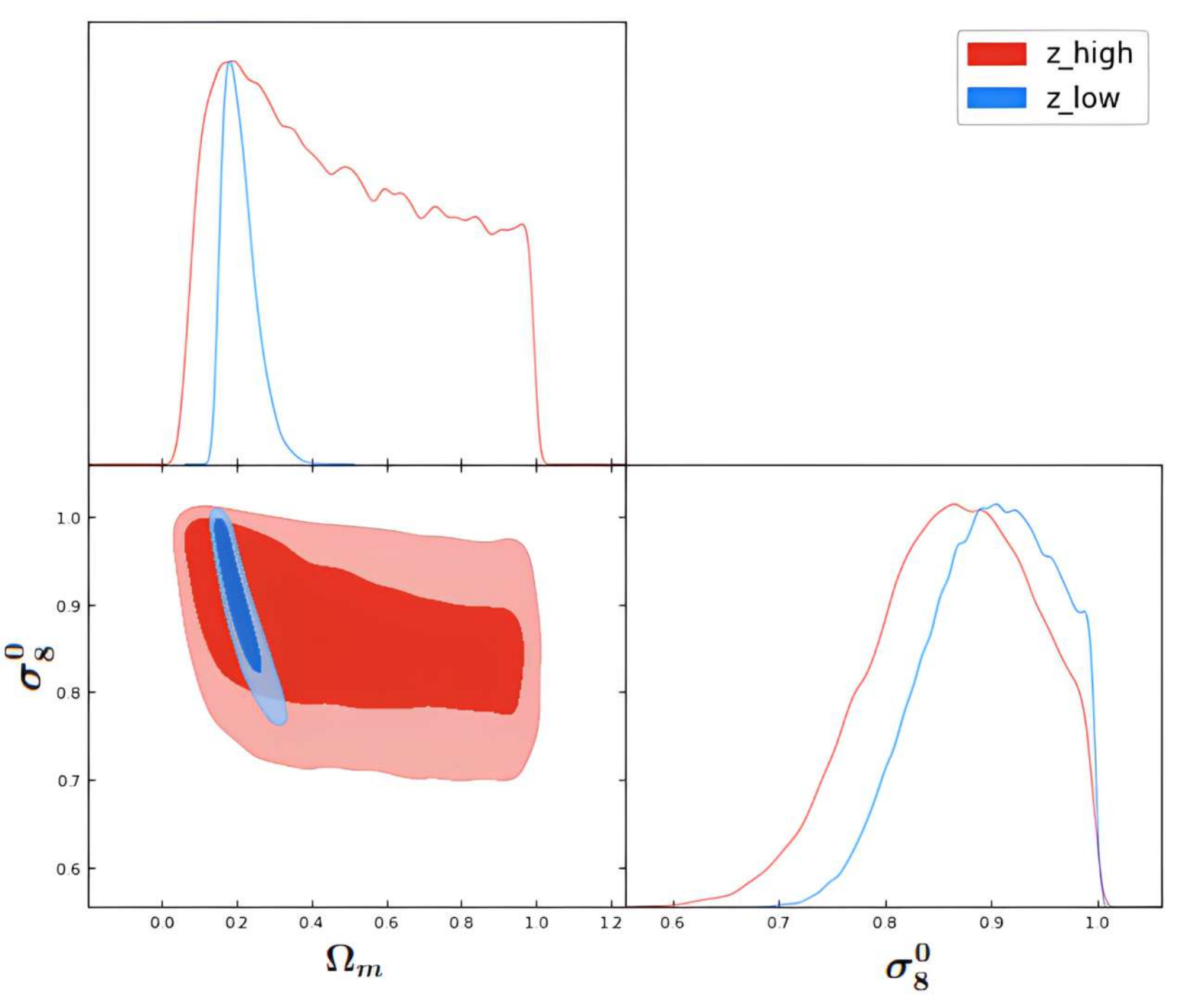}
  \caption{Marginalized credible intervals for $\Omega_m$ and $\sigma_8$ using 23 growth rate data points. The $f\sigma_8(z)$ data is bifurcated at $z = 0.7$.The innermost contour represents a 68\% credible interval, while the outermost contour corresponds to a 95\% credible interval. }
  \label{fig:Figure1}
\end{figure*}
\begin{figure*}[!t]
  \centering
  \includegraphics[width=0.8\textwidth]{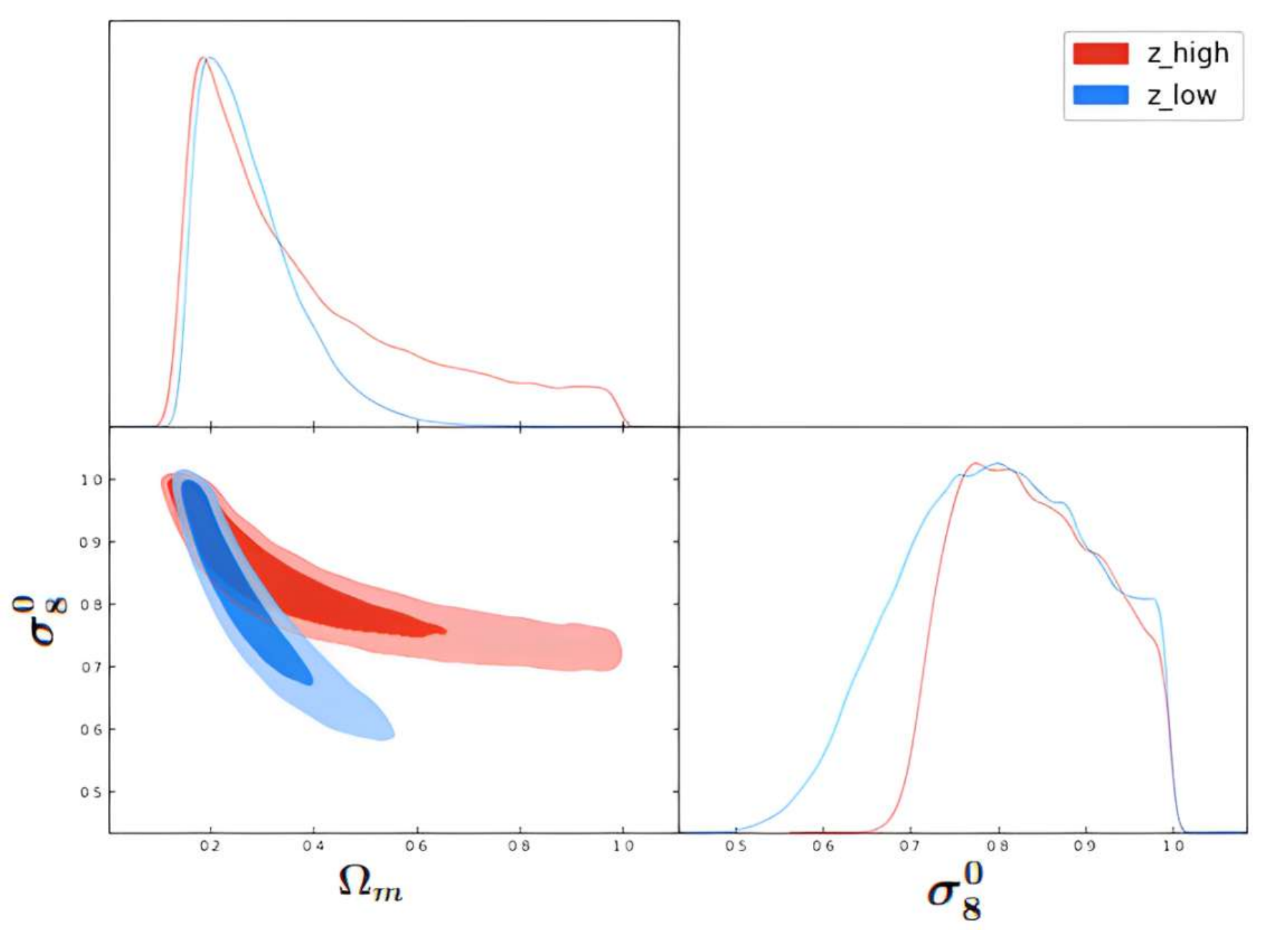}
  \caption{Same as Fig.~\ref{fig:Figure1} but $f\sigma_8(z)$ data is bifurcated at $z = 0.4$.}
  \label{fig:Figure2}
\end{figure*}
\begin{figure*}[!t]
  \centering
  \includegraphics[width=0.8\textwidth]{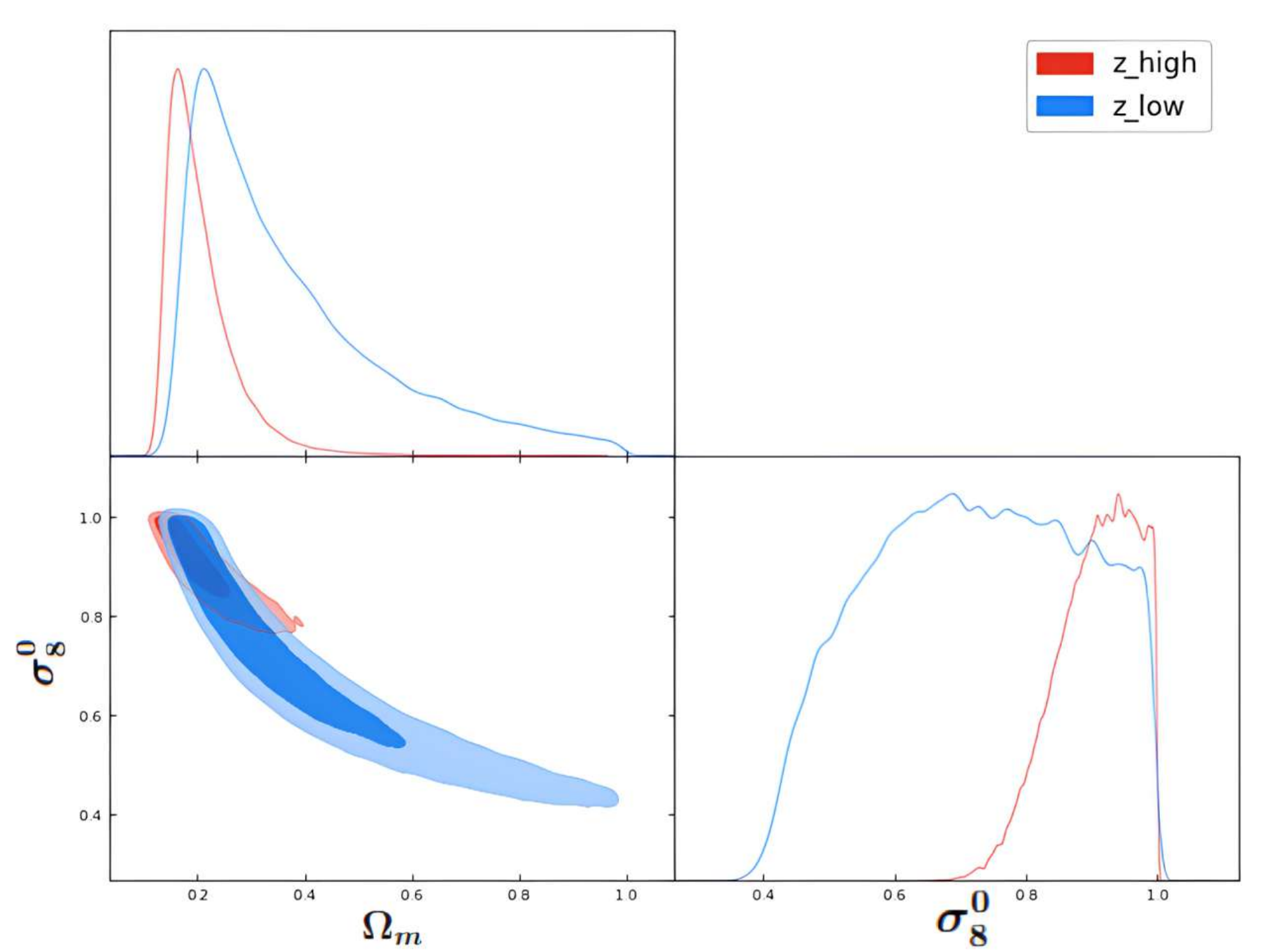}
  \caption{Same as Fig.~\ref{fig:Figure1} but $f\sigma_8(z)$ data is bifurcated at $z = 0.2$.}
  \label{fig:Figure3}
\end{figure*}

%The division of data at $z = 1.1$ in \ref{fig:Figure1} follows the methodology introduced in \cite{Adil2023}. 

\begin{table}[ht]
\centering
\begin{tabular}{|l|c|c|}
\hline
\textbf{Parameter} & \textbf{Distribution} & \textbf{Prior} \\ \hline
$\Omega_m$ & Uniform & $\mathcal{U}(0, 1.0)$ \\
$\sigma_8^0$ & Uniform & $\mathcal{U}(0.7, 0.9)$ \\
%$a$ & Uniform & $\mathcal{U}(0, 1)$ \\
%$b$ & Uniform & $\mathcal{U}(0, 1)$ \\
\hline
\end{tabular}
\caption{\textbf{Priors on the cosmological parameters used for our analyses.}}
\label{tab:Table1}
\end{table}

\begin{table}[ht]
\centering
\begin{tabular}{|l|c|c|c|}
\hline
\textbf{Data points} & \textbf{$\boldsymbol{\Omega_m}$} & \textbf{$\boldsymbol{\sigma_8^0}$} & \textbf{Redshift cut}\\ \hline
$07$ & $0.489 \pm 0.26$ & $0.863 \pm 0.075$ & $z>=0.7$\\
$16$ & $0.21 \pm 0.044$ & $0.898 \pm 0.059$ & $z<0.7$\\ \hline 
$12$ & $0.387 \pm 0.22$ & $0.841 \pm 0.078$ & $z>=0.4$\\
$11$ & $0.278 \pm 0.096$ & $0.803 \pm 0.105$ & $z<0.4$\\ \hline 
$16$ & $0.206 \pm 0.065$ & $0.906 \pm 0.061$ & $z>=0.2$\\
$07$ & $0.376 \pm 0.183$ & $0.723 \pm 0.154$ & $z<0.2$\\
\hline
\end{tabular}
\caption{\textbf{Best-fit values for $\sigma_8^0$ and $\Omega_m$  for all the four analyses carried out in this work.}}
\label{tab:Table2}
\end{table}

%\subsection{Discussion on Priors}
To investigate the redshift dependence, instead of carrying out regression using three free parameters ($\sigma_8^0$, $S_8$ and $\Omega_m$) as in A23, we use only two parameters,  $\sigma_8^0$ and $\Omega_m$.  We then use a Gaussian likelihood defined as follows:
\begin{equation}
\ln \mathcal{L}(\Omega_m) = -\frac{1}{2} \sum_i \frac{\left[f_{\sigma_8, z_{i}} - \hat{f}_{\sigma_8, z_{i}}\right]^2}{\sigma_{f_{\sigma_8, z_{i}}}^2} \label{eq:4}
\end{equation}
%\begin{equation}
%\ln \mathcal{L}(\Omega_m) = -\frac{1}{2} \sum_i \frac{\left[f \sigma_8 - \hat{f}_{\sigma_8, z_{ i}}\right]^2}{\sigma_{f_{\sigma_8,z}^2}},
%\end{equation}
where $f\sigma_8$ is defined in Eq.~\ref{eq:fsigma8}.
We mostly use similar priors as in A23. The only exception is  $\Omega_m$, where we used a uniform prior $\in [0,1]$. For $\sigma_8^0$, we use a uniform prior. A tabular summary of the parameters and their respective priors can be found in Table~\ref{tab:Table1}. We then sample the above likelihood using the {\tt emcee} MCMC sampler~\cite{emcee}. The marginalized credible intervals are obtained using the {\tt getdist} package~\cite{getdist}.

We carried out two different analyses. We first did a Bayesian regression on the above parameters after bifurcating the dataset into two redshift samples with cuts at $z=0.2$, $z=0.4$, and  $z=0.7$.  
%For the second analysis, we parameterize the redshift dependence of $\sigma_8^0$  according to $\sigma_8^0 \equiv (a + bz) $, where $a$ and $b$ are free parameters, similar to the models used to check for a redshift dependence of the gas depletion parameter in galaxy clusters~\cite{Bora}.  We plug this expression for $\sigma_8(z)$ instead of a constant $\sigma_8$ in Eq.~\ref{eq:fsigma8}. We impose uniform priors on both $a$ and $b$, restricting their values to the range between 0 and 1 (cf. Table~\ref{tab:Table1}).

%The Gaussian prior for the matter density parameter ($\Omega_m$) is informed by a comprehensive synthesis of constraints derived from the Cosmic Microwave Background (CMB), galaxy distributions, quasar properties, and Baryon Acoustic Oscillations (BAO) in the Lyman-$\alpha$ forest \cite{Planck2020,Alam2017,Hou2021,Neveux2020}. The mean and standard deviation of this Gaussian prior are set to $\mu_{\Omega_m} = 0.3111$ and $\sigma_{\Omega_m} = 0.0056$, respectively as per \cite{Adil2023}.

%For the fluctuation amplitude ($\sigma_8$), we adopt a uniform prior between 0.7 and 0.9, reflecting an equal probability for all values within this range.

 The best-fit parameters for $\Omega_m$ and $\sigma_8$ for 23  can be found in Table~\ref{tab:Table2}.
 The corresponding marginalized contours for both the low and high redshift samples can be found  in Fig.~\ref{fig:Figure1}.  
 %The corresponding contours for redshift dependent $\sigma_8^0$ can be found in Fig.~\ref{fig:Figure2}.  We also show the best fit for a redshift dependent $\sigma_8^0$ superposed on the data  in Fig.~\ref{fig:Figure3}. 
 We summarize our results  as follows:
\begin{itemize}
    \item The difference in $\Omega_m$ values between both the low redshift and high redshift sample for both sets of data is less than $1\sigma$, for all the three redshift cuts. Therefore, we do not find any evidence for a redshift  dependence of $\Omega_m$. 
    \item The difference in $\sigma_8$ values between the low redshift and high redshift sample is also less than $1\sigma$ for  our 23 measurements dataset for all the three redshift cuts.
   % \item Our best-fit values for $\sigma_8$ are consistent the Planck 2020 measurements for both the low-redshift and high-redshift samples.
    %\item When we model $\sigma_8^0$ as a linear function of redshift, we find $b=0.21^{+0.008}_{-0.18}$.   Therefore, \rthis{the redshift-dependent term is consistent with zero}  at 1.2$\sigma$.    
    %\item For a linear $\sigma_8^0(z)$, we find a reduced \rthis{ $\chi^2$ of 0.8.} 
   
\end{itemize}

Therefore, unlike A23, we do not find any dependence of $\sigma_8^0$ with redshift, and our results are  in accord with $\Lambda$CDM.

%\clearpage
\section{\label{sec:conclusions}Conclusions\protect}
To get some insight on the $\sigma_8$/$S_8$ tension problem, A23 looked for a redshift dependence of $\sigma_8^0$ and $S_8$ using $f\sigma_8$ measurements obtained from peculiar velocity and RSD measurements. They carried out a regression analysis using $\Omega_m$, $\sigma_8^0$, and $S_8$ and found a redshift dependence of $S_8$ with a discrepancy of $1.6-2.8\sigma$ between the low  and high redshift samples.

We carry out a variant of the above analysis using an independent dataset which has been thoroughly  vetted using internal consistency checks~\cite{Sagredo}, by doing  a regression analysis using only $\Omega_m$ and $\sigma_8^0$, and bifurcating the dataset into low and high redshift samples, after imposing a  redshift cut of $z=0.2$, $z=0.4$, and $z=0.7$.
We use a uniform prior on $\Omega_m$ and uniform prior on $\sigma_8^0$. We find that $\sigma_8^0$ values are consistent between the low redshift and high redshift samples to within $1\sigma$.
%When we parameterize the redshift dependence of $\sigma_8^0$ \rthis{using a linear dependence on redshift, we find the redshift-dependent term is consistent with 0 to within $1\sigma$}.
Therefore, we do not find evidence for an increase of $\sigma_8^0$ with redshift which was found in A23, without incorporating $S_8$ in our analysis. This implies that there is no breakdown of $\Lambda$CDM using our analysis, \rthis{assuming there are no uncontrolled systematics in the dataset hitherto analyzed.}
%Further confirmations can be obtained using upcoming data from spectroscopic surveys such as DESI~\cite{DESI}.

\begin{acknowledgments}
SM extend his sincere gratitude to the Government of India, Ministry of Education (MOE) for their continuous support through the stipend, which has played a crucial role in the successful completion of our research. We are also grateful to  Eoin Colgain, Yvonne Wong and  the anonymous referee  for comments and constructive feedback on the manuscript.

\end{acknowledgments}

%\clearpage
%\bibliographystyle{apsrev4-1} % Choose a bibliography style; here, we're using the APS style
%\newpage
%\clearpage
\bibliography{references}

%\bibliography{references}
\end{document}